\documentclass[useAMS,usenatbib]{mn2e}

\usepackage[fleqn]{amsmath}
\usepackage{amssymb}
\usepackage{graphicx}

\newcommand{\bfx}[1]{#1}

\newcommand{\Cklj}{C_{{\bf j}}^{{\bf l}}}
\newcommand{\Ham}{{\cal H}}

\newcommand{\e}{{\rm e}}

\def\abs#1{\left\vert#1\right\vert}

\newcommand{\moy}[2]{\left\langle{#2}\right\rangle_{#1}}

\def\crm{\cr\noalign{\medskip}}

\def\m@th{\mathsurround=0pt}
\def\EQM#1{\vcenter{\normalbaselines\m@th
    \ialign{${\displaystyle ##}$\hfil&&\ ${\displaystyle ##}$\hfil\crcr
    \mathstrut\crcr\noalign{\kern-\baselineskip}
    \noalign{\smallskip}
    #1\crcr\mathstrut\crcr\noalign{\kern-\baselineskip}}}}

\newcommand{\be}{\begin{equation}}
\newcommand{\ee}{\end{equation}}

\def\Dron#1#2{\frac{\partial#1}{\partial#2}}

\newcommand{\bpm}{\begin{pmatrix}}
\newcommand{\epm}{\end{pmatrix}}
\newcommand{\figampl}{
\begin{figure}
\begin{center}
\includegraphics[width=0.9\linewidth, viewport=0 0 220 163, clip]{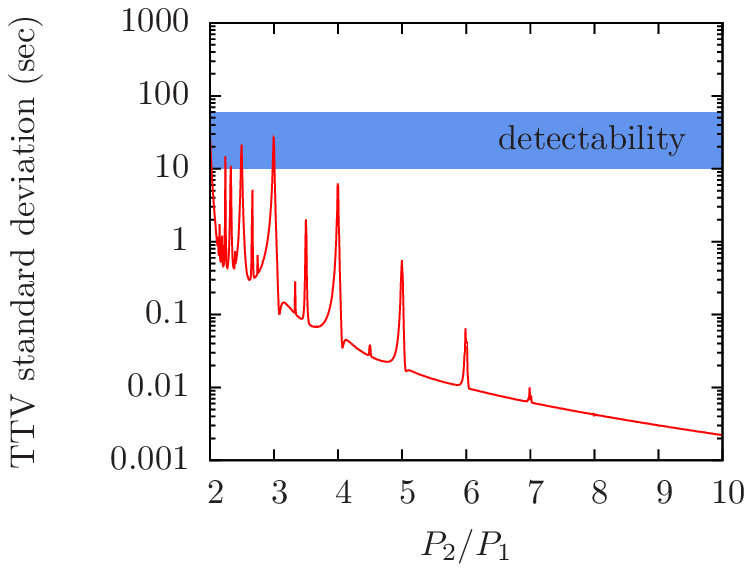}
\caption{\label{fig.ampl} Standard deviation of TTVs produced by an
Earth mass planet on a Jovian planet transiting a Solar mass star every
three days. The stripe shows the range of typical detectability
thresholds.}
\end{center}
\end{figure}
}

\newcommand{\figsignal}{
\begin{figure}
\begin{center}
\includegraphics[width=0.9\linewidth, viewport=0 0 270 380, clip]{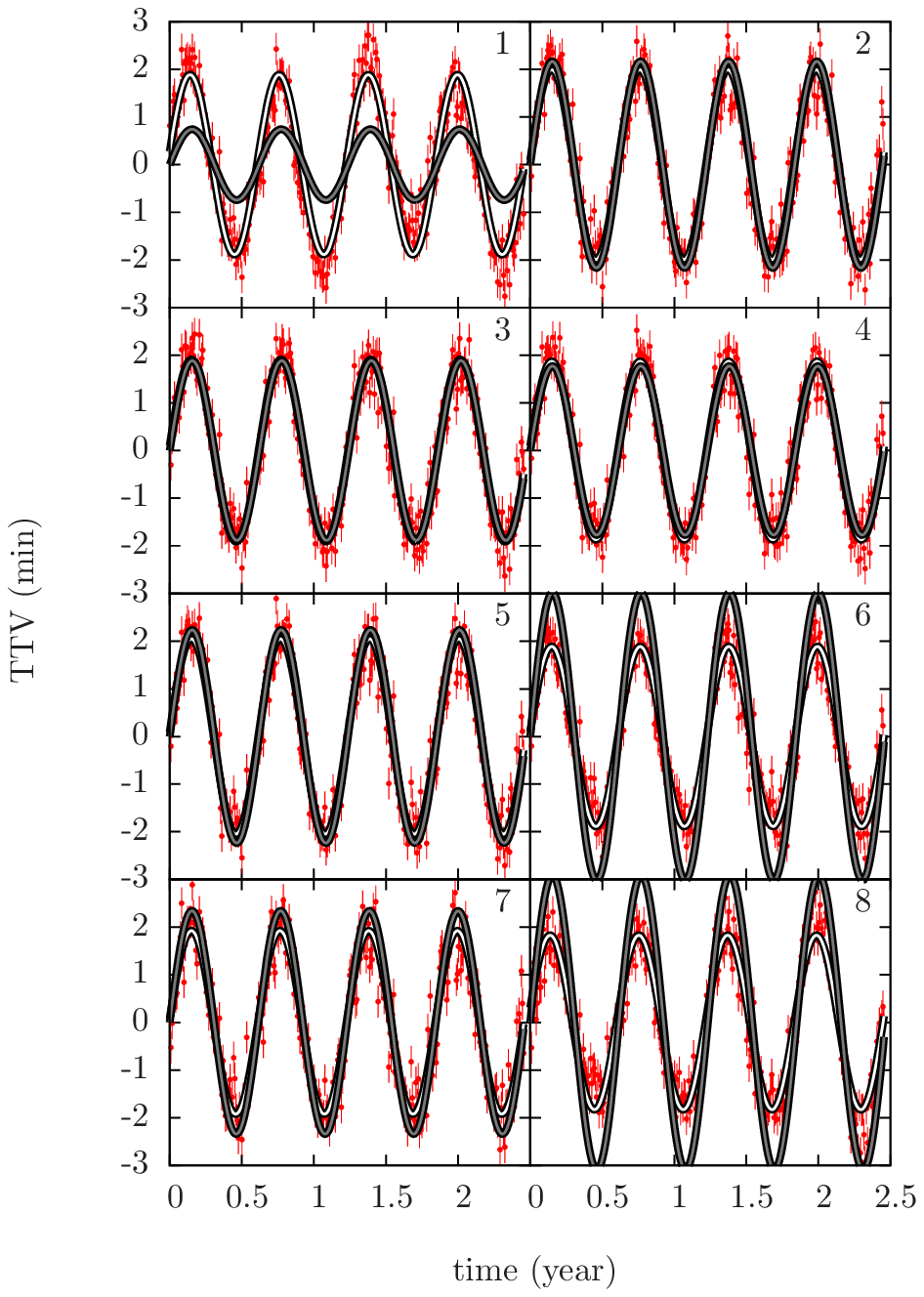}
\caption{\label{fig.signal} Example of TTV signal produced near period
commensurabilities. A gaussian error of 20 seconds has been added to the
signal to model the observation uncertainties. The white curve represents
the highest amplitude term obtained by frequency analysis. The gray
curve is the analytic approximation Eq.~(\ref{eq.deltat1}). The number
at the upper right corner of each subfigure corresponds to the set
of initial conditions as given in table~\ref{tab.param}.}
\end{center}
\end{figure}
}

\newcommand{\figcrossing}{
\begin{figure}
\begin{center}
\includegraphics[width=\linewidth, viewport= 0 0 220 130, clip]{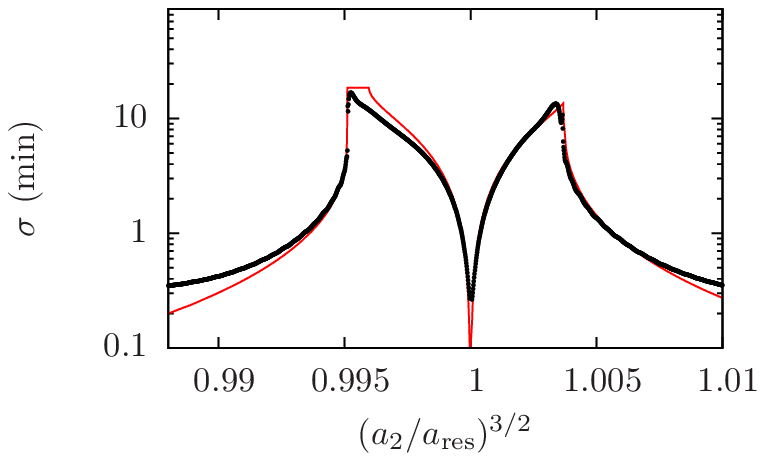}
\caption{\label{fig.crossing} Detail of the TTV amplitude through the
$3:7$ MMR. The initial conditions are those of the set 3
(Tab.~\ref{tab.param}) with the initial mean longitude
$\lambda_2=90\deg$. Each black point corresponds to a different initial
semi-major axis $a_2$ of the perturber. The thin solid line corresponds
to the analytical formulae (\ref{eq.deltat1}), (\ref{eq.deltat2})
taking into account the amplitude of libration $\Delta\psi$ inside the resonance. 
\bfx{The amplitude of libration is computed from
the initial conditions using the relation $\tilde \Ham(\psi(0),\Psi(0))
= \tilde \Ham(\psi_{\rm res} \pm \Delta\psi,\Psi_{\rm res})$ where $\tilde
\Ham$ is the resonant Hamiltonian (\ref{eq.Hamtilde2}).}}
\end{center}
\end{figure}
}

\newcommand{\tabparam}{
\begin{table}
\begin{center}
\caption{\label{tab.param} Planetary parameters used to perform the
simulations of the Figure~\ref{fig.signal}. }
\begin{tabular*}{\linewidth}{@{\extracolsep{\fill}}crrrrr} 
\hline\hline
set$^{\rm a}$ & MMR & $m_2$ ($M_\oplus$) & $e_2{}^{\rm b}$ & $\varepsilon$ & $\sqrt{\chi_r^2}$ \\ 
\hline\hline
1  & $1:2$     &   0.9 & 0.087 & 0.0132 & 1.36 \\
2  & $4:9$     &  24.5 & 0.120 & 0.0034 & 1.15 \\
3  & $3:7$     &  21.1 & 0.100 & 0.0044 & 1.13 \\
4  & $2:5$     &   8.6 & 0.102 & 0.0067 & 1.02 \\
5  & $3:8$     &  17.2 & 0.160 & 0.0045 & 1.07 \\
6  & $1:3$     &  49.7 & 0.100 & 0.0134 & 1.25 \\
7  & $3:10$    &  95.2 & 0.194 & 0.0045 & 1.30 \\
8  & $1:4$     & 394.0 & 0.115 & 0.0134 & 1.31 \\ \hline
\end{tabular*}
\end{center}
\bfx{
$^{\rm a}$ The initial conditions
are in order of increasing semi-major axis. \\
$^{\rm b}$ The eccentricity $e_2$ is chosen to produce low values of the
reduced chi square $\sqrt{\chi_r^2}$.}
\end{table}
}

\title[Degeneracy of the transit timing variation method]%
{Degeneracy in the characterization of
non-transiting planets from transit timing variations}
\author[G. Bou\'e et al.]{G. Bou\'e$^{1,2}$\thanks{E-mail:
gwenael.boue@astro.up.pt}, M. Oshagh$^{1,3}$, M. Montalto$^{1}$, and N. C.
Santos$^{1,3}$\\
$^{1}$Centro de Astrof\'isica, Universidade do Porto, 
              Rua das Estrelas, 4150-762 Porto, Portugal \\
$^{2}$ASD, IMCCE-CNRS UMR8028, Observatoire de Paris, UPMC, 77
             avenue Denfert-Rochereau, 75014 Paris, France \\
$^{3}$Departamento de F\'isica e Astronomia, Faculdade de
             Ci\^encias, Universidade do Porto, Portugal}
\begin{document}

\date{Accepted \ldots. Received \ldots; in original form \ldots}

\pagerange{\pageref{firstpage}--\pageref{lastpage}} \pubyear{\ldots}

\maketitle

\label{firstpage}
 
\begin{abstract}
   The transit timing variation (TTV) method allows the detection of
non-transiting planets through their gravitational perturbations.  Since
TTVs are strongly enhanced in systems close to mean-motion resonances
(MMR), even a low mass planet can produce an observable signal. This
technique has thus been proposed to detect terrestrial planets.  In this
letter, we analyse TTV signals for systems in or close to MMR in order
to illustrate the difficulties arising in the determination of planetary
parameters.  TTVs are computed numerically with an $n$-body integrator
for a variety of systems close to MMR. The main features of these TTVs
are also derived analytically.  Systems deeply inside MMR do not produce
particularly strong TTVs, while those close to MMR generate
quasiperiodic TTVs characterised by a dominant long period term and a
low amplitude remainder. If the remainder is too weak to be detected,
then the signal is strongly degenerate and this prevents the
determination of the planetary parameters.  Even though an Earth mass
planet can be detected by the TTV method if it is close to a MMR, it may
not be possible to assert that this planet is actually an Earth mass
planet.  On the other hand, if the system is right in the center of a
MMR, the high amplitude oscillation of the TTV signal vanishes and the
detection of the perturber becomes as difficult as it is far from MMR.
\end{abstract}


%
%________________________________________________________________

\section{Introduction}
A transiting planet belonging to a multiplanetary system may suffer
gravitational interactions. The perbutations in the orbit can then
produce observable variations in the transit midtimes with respect to
those of a purely Keplerian orbit. This is the basic idea behind the
transit timing variation method which has been proposed to detect
non-transiting terrestrial planets \citep{Agol_etal_MNRAS_2005,
Holman_Murray_science_2005}.

The TTV signal scales roughly linearly with the mass of the perturber
\bfx{\citep{Agol_etal_MNRAS_2005, Holman_Murray_science_2005,
Nesvorny_Morbidelli_ApJ_2008}}. Thus, low-mass planets generate weak
TTVs except if they are in mean motion resonance or near period
commensurability with the transiting planet \citep{Agol_etal_MNRAS_2005,
Holman_Murray_science_2005}.  In that case, the TTV signal \bfx{contains
a long term oscillation associated to a ``great inequality'', as for
the Jupiter-Saturn system
\citep{Lissauer_etal_nature_2011,Ballard_etal_ApJ_2011}. This long
period, like all the oscillations in the TTV signal, scales linearly
with the period of the transiting planet and is of the order of a year
for a three-day transiting planet.} Such systems, if they exist, should
thus start to be detectable in surveys like CoRoT or Kepler. However,
most of the systems close to MMR generate similar TTVs
\bfx{\citep{Ford_Holman_ApJL_2007}}, and  many perturber parameters can 
reproduce an observed TTV signal \citep{Nesvorny_Morbidelli_ApJ_2008,
Meschiari_Laughlin_ApJ_2010, Veras_etal_ApJ_2011}. Such a difficulty has
already been experienced with the Kepler 19 system
\citep{Ballard_etal_ApJ_2011}. \bfx{Nevertheless, short-term
perturbations, if detectable, should produce a ``chopping signal'' that
raises the degeneracy \citep{Holman_etal_science_2010}.}

Analytical expressions of the order of magnitude of the amplitude and
the period of TTV signals for systems in or near first-order mean-motion
resonances are provided by \citet{Agol_etal_MNRAS_2005}. Here, we propose
to focus on higher order resonances. In a first section, we compare
the TTV signal produced by an Earth mass planet with typical detection
thresholds to motivate the study of these resonances. In a second
section, we present a few \bfx{examples} of TTV signals for a variety of
planetary parameters. In the subsequent section, we provide analytical 
approximations of the TTV signal for systems near or at MMR. 
We conclude in the last section.

\section{TTV amplitude produced by an Earth mass planet}
\label{sec.earth}
We consider the strength of TTVs produced by a terrestrial planet of
mass $m_2=M_\oplus$ on a Jupiter mass planet ($m_1=M_{\rm J}$)
transiting a star with mass $m_0=M_\odot$. The orbit of the transiting
planet is assumed to be \bfx{initially} circular with a period $P_1=3$ days.
Since second and higher order mean motion resonances only exist for non
zero eccentricities, we set the eccentricity of the perturbing planet to
$e_2=0.1$. The TTVs are computed using an $n$-body integrator over
$N=300$ transits which corresponds to 2.5 years. The system is supposed
to be coplanar, the initial mean longitudes $\lambda_1$, $\lambda_2$ of
the two planets and the longitude of the periapsis $\varpi_2$ of the
perturber are \bfx{arbitrarily} set to $270\deg$ which is the direction
of the observer.

The strength of TTVs is measured in the same way as in
\bfx{\citet{Agol_etal_MNRAS_2005}}, using the standard deviation
\be
\sigma = \sqrt{\frac{1}{N}\sum_{j=1}^{N}
\left(t_j-t_0+P_1j\right)^2}\ ,
\label{eq.sigma}
\ee
where $P_1$ and $t_0$ are chosen to minimize $\sigma$, and $t_j$, 
$j \geq 1$, are the midtransit times of the $N$ transits.
This quantity $\sigma$ is displayed in figure~\ref{fig.ampl} for 1000 different
initial conditions evenly spread over the period $P_2$ of the perturber
between $2P_1$ and $10P_1$. We recognize the well known \bfx{peaks}
centered on period commensurabilities corresponding to amplified
gravitational interactions \citep{Agol_etal_MNRAS_2005,
Holman_Murray_science_2005}.  The figure shows also the range of typical
detection thresholds varying from 10 sec to 1 min depending on the
instrument and on the depth of the transit. It is noteworthy that, for
the system considered in this letter, terrestrial planets can hardly
produce signals at the limit of detection and only in the vicinity of
period commensurabilities with $P_2/P_1 \leq 3$. \bfx{Since the
amplitude of TTV scales linearly with the period of the
transiting planet, the signal produced on a wider orbit, with a period
ten times larger for instance, would be easier
to detect. However, the period of the TTV signal would be ten
times longer too. It may exceed the duration of the surveys like CoRoT
and Kepler.}
% More massive planets may generate detectable signals at larger orbital
% periods. 
With this choice of initial conditions, most of the systems
do not enter MMR because the value of the resonant angle is that of the
hyperbolic point (see section~\ref{sec.immr}).

\figampl

\section{TTV signal near a period commensurability}
\label{sec.simu}
To illustrate the degeneracy of the TTV signal, we consider 8
different initial conditions, near different mean motion resonances. The
system is the same as in the previous section, \bfx{apart} from the mass of the
perturber $m_2$ and its eccentricity $e_2$ which are chosen so as to
produce TTVs with the same amplitude ($\sim 2$ min) and the same period (4
oscillations during the 2.5 years) for each of the different orbital periods
$P_2$. The initial conditions are gathered in Table~\ref{tab.param}. 
Although this letter focuses on second and higher order period
commensurabilities, we consider the 1:2 resonance to highlight the fact
that even a 0.9 Earth mass planet can generate a TTV signal with the
same amplitude and shape as a 4.14 Saturnian mass planet near the 1:4
resonance.

\tabparam

Figure~\ref{fig.signal} shows the TTV signals, computed for the 8
initial conditions of table~\ref{tab.param}, to which a gaussian noise
with $\sigma_{\rm noise}=20$ seconds has been added to simulate
observations. All the signals are very similar and easily detectable.
But they are dominated by one single frequency and look like simple
noisy sinusoids. It is thus quite difficult to differentiate them. 
For each signal, the highest amplitude sinusoid is extracted using a
frequency analysis \citep{Laskar_Icarus_1990, Laskar_PhysD_1993}, and
represented by a white curve in Fig.~\ref{fig.signal}. Then, the reduced
chi-square of this fit is computed and shown in the last column of the
table~\ref{tab.param}.  In all cases, the fit by a pure sine leads to a
reduced chi-square lower than or of the order of 1.30.
\bfx{Naturally, if the two planets transit, the degeneracy disappears
because the resonance is known. Here, we focus only on non-transiting
perturbers.}

In the following we derive an analytical approximation of this highest
amplitude sinusoidal term.

\figsignal

\section{Analytical approximation of the highest amplitude term}
\label{eq.analytic}
We consider a coplanar system in the vicinity of the $p:p+q$ mean motion
resonance. We assume that the orbit of the transiting planet is
initially circular, while that of the perturber is eccentric.
We define $a_i$, $\lambda_i$, $\e_i$, $\varpi_i$ as the semi-major axis,
the mean longitude, the eccentricity and the longitude of pericenter of
the $i$th planet, respectively. The conjugated variable to the mean
longitude $\lambda_i$ is, up to the first order in planet masses, 
$\Lambda_i=m_i\sqrt{G m_0 a_i}$ with $G$ being the gravitational
constant. The index $i=1$ stands for the inner (transiting) planet, and
the index $i=2$ refers to the outer perturber.  For each orbital
parameter $E_i$, we note the unperturbed part $\bar E_i$, and the
oscillatory perturbation $\delta E_i$ such that $E_i = \bar E_i + \delta
E_i$.  For the system analysed in this section, the terms contributing
to the transit timing variations $\delta t$ of the inner planet are
\citep[e.g.][]{Nesvorny_Morbidelli_ApJ_2008}
\be
n_1 \delta t = \delta \lambda_1 + 2 \delta k_1 \sin(\bar \lambda_1)
- 2\delta h_1 \cos(\bar \lambda_1) + {\cal O}(e_1)\ .
\label{eq.deltat0}
\ee
In this expression, $n_1=2\pi/P_1$, defined by $Gm_0=n_1^2 a_1^3$, is
the mean motion of the transiting planet. $k_1 = e_1 \cos \varpi_1$ and
$h_1 = e_1 \sin \varpi_1$ are the usual non-singular eccentricity
variables of the transiting planet. With the convention $\bar
\lambda_1=3\pi/2$ at transit, Eq.~(\ref{eq.deltat0}) becomes
\be
n_1 \delta t = \delta \lambda_1 -2 \delta k_1 + {\cal O}(e_1)\ .
\label{eq.deltat}
\ee

The Hamiltonian $\Ham$ governing the evolution of a coplanar two-body
system can be decomposed into a Keplerian part 
\be
\Ham_0 = -\frac{G m_0 m_1}{2 a_1}
-\frac{G m_0 m_2}{2 a_2}\ ,
\label{eq.Ham0}
\ee
and an interaction $\Ham_1$ given by
\be
\EQM{
\Ham_1 
&=& -\frac{G m_1 m_2}{a_2} \sum_{{\bf j}, {\bf l}} \Cklj(\alpha)
e_1^{l_1} e_2^{l_2} \crm && \times
\cos\big(j_1 \lambda_1 + j_2 \lambda_2 + j_3 \varpi_1 + j_4
\varpi_2\big)\ ,
}
\label{eq.Ham1}
\ee
where $\alpha=a_1/a_2$. $\Cklj(\alpha)$ with ${\bf
j}=(j_1,j_2,j_3,j_4)$ and ${\bf l}=(l_1,l_2)$ are coefficients that can
be computed either in terms of Laplace coefficients ${\rm
b}_{s/2}^{(k)}(\alpha)$ \citep[e.g.][]{Ellis_Murray_Icarus_2000}, or
through a series in power of $\alpha$ \citep[e.g.][]{Kaula_AJ_1962}.

Here, we make the assumption that the evolution of a system in the
vicinity of the $p:p+q$ MMR is well recovered when only the resonant
terms with the lowest power in eccentricity are kept in
Eq.~(\ref{eq.Ham1}). This is verified if the resonances are well
separated \citep{Wisdom_AJ_1980}. One gets
\be
\EQM{
\Ham_1 &\simeq&
 -  e_2^q H_1 \cos
\big(p \lambda_1 - (p+q) \lambda_2 + q \varpi_2\big) \crm
& -&  e_1 e_2^{q-1} H_2 \cos
\big(p \lambda_1 - (p+q) \lambda_2 + \varpi_1 + (q\!-\!1) \varpi_2\big)
}
\label{eq.Ham}
\ee
where $H_i = 2 G m_1 m_2 C_i(\alpha) / a_2$ and
\be
C_1(\alpha) = C_{(p, -p-q, 0, q)}^{(0,q)}
(\alpha)\ ,\ 
C_2(\alpha) = C_{(p, -p-q, 1, q-1)}^{(1,q-1)}
(\alpha)\ .
\ee
The factor 2 in the definition of $H_i$ comes from the symmetry
$\Cklj(\alpha) = C_{-{\bf j}}^{{\bf l}}(\alpha)$.
In (\ref{eq.Ham}), we consider only the linear terms in $e_1$ as the
higher orders do not contribute to the equations of motion for $\bar
e_1=0$.  Since the eccentricity $e_2$ does not enter in the expression
of $\delta t$ (\ref{eq.deltat}), we make the hypothesis that
$e_2$ and its longitude of pericenter $\varpi_2$ are dominated by their
constant (initial) value for the period of time considered in this
study. Thus, $e_2$ and $\varpi_2$ are not variables, but just parameters
of the problem. Hereafter, we note them $\bar e_2$ and $\bar \varpi_2$,
respectively.

The evolution dictated by the resonant Hamiltonian, Eqs.~(\ref{eq.Ham0})
and (\ref{eq.Ham}), is obtained after applying the following conical
transformation on the mean longitudes
\be
\EQM{
\phi &=& -\lambda_1/(p+q) \ , \crm
\psi &=& p \lambda_1 - (p+q) \lambda_2 + q \bar \varpi_2 \ ,
}
\label{eq.angle}
\ee
and on their conjugated momenta
\be
\EQM{
\Phi &=& -(p+q) \Lambda_1 - p \Lambda_2 \ , \crm
\Psi &=& - \Lambda_2 / (p+q) \ .
\label{eq.action}
}
\ee
In the new variables, the Hamiltonian reads
\be
\EQM{
\tilde \Ham  & (\psi, \Phi, \Psi, h_1, k_1) =
\tilde \Ham_0(\Phi, \Psi) - \bar e_2^q \tilde H_1(\Phi,\Psi) \cos \psi
\crm &
- \bar e_2^{q-1} \tilde H_2(\Phi,\Psi) \big(k_1 \cos (\psi-\bar \varpi_2)
- h_1 \sin (\psi-\bar \varpi_2)\big) \ .
}
\label{eq.Hamtilde}
\ee
It should be stressed that $k_1$ and $h_1$ are not conical variables.
% The equations of motion are
% \be
% \EQM{
% \dot \phi &=& \Dron{\tilde \Ham}{\Phi} -
% \frac{1}{p+q} \frac{\beta_1}{\Lambda_1(1+\beta_1)}\left(
% k_1 \Dron{\tilde \Ham}{k_1}+h_1\Dron{\tilde \Ham}{h_1}\right)\ ,
% \crm 
% \dot \psi &=& \Dron{\tilde \Ham}{\Psi} + p
% \frac{\beta_1}{\Lambda_1(1+\beta_1)}\left(
% k_1 \Dron{\tilde \Ham}{k_1}+h_1\Dron{\tilde \Ham}{h_1}\right)\ ,
% \crm 
% \dot \Phi &=& -\Dron{\tilde \Ham}{\phi}\ , \crm
% \dot \Psi &=& -\Dron{\tilde \Ham}{\psi}\ , \crm
% \dot k_1  &=& \frac{\beta_1}{\Lambda_1}\left(\Dron{\tilde \Ham}{h_1}
% +\frac{k_1}{1+\beta_1}\left(-\frac{1}{p+q} \Dron{\Ham}{\phi} + 
% p \Dron{\Ham}{\psi}\right)\right)\ ,
% \crm
% \dot h_1  &=& \frac{\beta_1}{\Lambda_1}\left(-\Dron{\tilde \Ham}{k_1}
% +\frac{h_1}{1+\beta_1}\left(-\frac{1}{p+q} \Dron{\Ham}{\phi} + 
% p \Dron{\Ham}{\psi}\right)\right)\ ,
% }
% \label{eq.motion1}
% \ee
% with $\beta_1=\sqrt{1-e_1^2}$. 
However, in the limit of vanishing eccentricity $e_1$, the equations of
motions are greatly simplified
\be
\EQM{
\dot \phi =  \Dron{\tilde \Ham}{\Phi}\ ,\quad&
\dot \psi =  \Dron{\tilde \Ham}{\Psi}\ ,\quad&
\dot k_1  = \frac{1}{\Lambda_1}\Dron{\tilde \Ham}{h_1}
\ , \crm
\dot \Phi = -\Dron{\tilde \Ham}{\phi}\ ,\quad&
\dot \Psi = -\Dron{\tilde \Ham}{\psi}\ ,\quad&
\dot h_1  = -\frac{1}{\Lambda_1}\Dron{\tilde \Ham}{k_1}
\ .
\label{eq.motion}
}
\ee
Since the Hamiltonian (\ref{eq.Hamtilde}) is independent of the cyclic
variable $\phi$, its conjugated momentum $\Phi$ is constant. In the 
following, we assume, as in \citet{Wisdom_AJ_1980}, that the functions
$\tilde H_i(\Phi,\Psi)$, $i=1,2$ of the interaction part are well
approximated by $\tilde H_i(\Phi,\bar \Psi)$, where $\bar \Psi = 
\moy{}{\Psi}$ is the averaged value of $\Psi$.

\subsection{Outside of MMR}
\label{sec.ommr}
Outside of the $p:p+q$ MMR, we assume that the Keplerian Hamiltonian 
$\tilde \Ham_0(\Phi, \Psi)$ is sufficiently well approximated by the 
linear term in its Taylor series about $\Psi=\bar \Psi$. In that case,
the resonant Hamiltonian (\ref{eq.Hamtilde}) becomes 
\be
\EQM{
\tilde \Ham &\simeq \tilde \Ham_0(\Phi, \bar \Psi) + 
\left.\Dron{\tilde \Ham_0}{\Psi}\right|_{\Psi=\bar \Psi} \!\!(\Psi\!-\!\bar \Psi)
-\bar e_2^q \tilde H_1(\Phi,\bar \Psi) \cos \psi \crm &
-\bar e_2^{q-1} \tilde H_2(\Phi,\bar \Psi) \big( 
k_1 \cos(\psi - \bar \varpi_2) - h_1 \sin(\psi - \bar \varpi_2)\big)\ ,
}
\label{eq.Hamtilde1}
\ee
with,
\be
\tilde H_i(\Phi,\bar \Psi) = 2 \bar \Lambda_1 n_1 \frac{m_2}{m_0} \bar \alpha
C_i(\bar \alpha)\ ,
\label{eq.Hi}
\ee
for $i=1,2$, and
\be
\left.\Dron{\tilde \Ham_0}{\Psi}\right|_{\Psi=\bar \Psi} \equiv
\omega_\psi(\Phi,\bar \Psi) = p n_1 - (p+q)
n_2 = \pm p n_1 \varepsilon\ .
\label{eq.H0lin}
\ee
In (\ref{eq.H0lin}), the parameter $\varepsilon$, which cancels out at
the resonance, is a measure of the closeness to the exact resonance. Its
expression is
\be
\varepsilon = \abs{1-\frac{p+q}{p}\frac{n_2}{n_1}}\ .
\ee

For infinitely small eccentricity $e_1$, the evolution of $\psi$ and
$\Psi$ governed by the Hamiltonian (\ref{eq.Hamtilde1}) is 
independent of $k_1$ and $h_1$. Using (\ref{eq.motion}), one finds
\be
\EQM{
\psi(t) = \psi(0) + \omega_\psi(\Phi,\bar \Psi) t\ , \crm
\Psi(t) = \Psi(0) + \frac{\tilde H_1(\Phi,\bar \Psi)}{
\omega_\psi(\Phi,\bar \Psi)} 
\bar e_2^q \big(\cos \psi(t) - \cos \psi(0) \big)\ .
}
\ee
To get the evolution of $\phi$, we assume that $\partial \tilde
\Ham/\partial \Phi$ is well approximated by the linear expansion 
of $\partial \tilde \Ham_0/\partial \Phi$ about $\Psi=\bar \Psi$, whose
the expression reads
\be
\Dron{\tilde \Ham_0}{\Phi} = -\frac{n_1}{p+q} + 3\frac{pn_1}{p+q}
\frac{\Psi-\bar \Psi}{\bar \Lambda_1} \ .
\label{eq.dphi}
\ee
Integrating (\ref{eq.dphi}) with $\bar \Psi = \moy{}{\Psi(t)}$, one gets
\be
\EQM{
\phi(t) =& \phi(0) - \frac{n_1 t}{p+q} \crm &+ 3\frac{p n_1}{p+q} \frac{\tilde
H_1(\Phi,\bar \Psi)}{\omega_\psi(\Phi,\bar \Psi)^2} \frac{\bar
e_2^q}{\bar \Lambda_1} 
 \big(\sin \psi(t) - \sin(\psi(0))\big)\ .
}
\label{eq.phit}
\ee
From the equations of change of variables (\ref{eq.angle}), using
the expression of $\tilde H_1(\Phi,\bar \Psi)$, Eq.~(\ref{eq.Hi}),
$\omega_\psi(\Phi,\bar \Psi)$, Eq.~(\ref{eq.H0lin}), and removing the
linear terms in (\ref{eq.phit}), one obtains
\be
\delta \lambda_1 = -\frac{6}{p} \frac{m_2}{m_0} \frac{\bar \alpha
C_1(\bar \alpha)}{\varepsilon^2} \bar e_2^q \sin \psi(t)\ .
\label{eq.dlambda1}
\ee
The expression of $\delta k_1$ is obtained from the integration of the
equations of motion (\ref{eq.motion}) of the Hamiltonian (\ref{eq.Hamtilde1}).
After removing the linear terms, the result is
\be
\delta k_1 = \mp \frac{2}{p} \frac{m_2}{m_0} \frac{\bar \alpha
C_2(\bar \alpha)}{\varepsilon} \bar e_2^{q-1} \cos (\psi(t)-\bar \varpi_2)\ .
\label{eq.dk1}
\ee
From the expressions (\ref{eq.dlambda1}) and (\ref{eq.dk1}), one can see
that the ratio between the amplitudes of $\delta \lambda_1$ and $\delta
k_1$ is of the order of $\bar e_2/\varepsilon$. Thus, for a system
sufficiently close to a MMR, $\varepsilon \lesssim \bar e_2$, the amplitude
of the TTV signal (\ref{eq.deltat}) is dominated by $\delta \lambda_1$.
In that case,
\be
\delta t \simeq -\frac{6}{pn_1} \frac{m_2}{m_0} \frac{\bar \alpha
C_1(\bar \alpha)}{\epsilon^2} \bar e_2^q \sin \psi(t)\ .
\label{eq.deltat1}
\ee
The period of the dominant sinusoidal oscillation in the TTV signal 
$P = 2\pi/\abs{\omega_\psi(\Phi,\bar \Psi)}$ is related to the period of
the transiting planet by (see Eq.~(\ref{eq.H0lin}))
\be
P = \frac{1}{p\varepsilon} P_1\ .
\label{eq.PP1}
\ee
This relation has been inverted to compute the distance $\varepsilon$
from each resonance in the different simulations of the
section~\ref{sec.simu}, see table~\ref{tab.param}.

The analytical approximation of TTVs (\ref{eq.deltat1}) is plotted as
gray curves for each numerical experiment in Fig.~\ref{fig.signal}. The
agreement with the numerical simulations is very good except for the
largest distances to MMR: $\varepsilon\sim 1.30$. Nevertheless, the
order of magnitude of the amplitude is still correct within a factor 2.

\subsection{Inside of MMR}
\label{sec.immr}
The reasoning is the same as in the case outside of MMR. The
only difference is that the linear terms in the Taylor series of $\tilde
\Ham_0(\Phi,\Psi)$ about $\Psi=\bar \Psi=\Psi_{\rm res}$ is zero. Thus, we
expand $\Ham_0(\Phi,\Psi)$ up to the quadratic order as in
\citet{Wisdom_AJ_1980}. Furthermore, we neglect the contribution 
$\delta h_1$ in the TTV, as in the previous subsection. The part
proportional to $\tilde H_2(\Phi,\Psi_{\rm res})$ is thus removed from
the Hamiltonian.  We get
\be
\EQM{
\tilde \Ham =& \tilde \Ham_0(\Phi,\Psi_{\rm res}) +
\frac{1}{2}\left.\Dron{^2\Ham_0}{\Psi^2}\right|_{\Psi=\Psi_{\rm res}}
(\Psi-\Psi_{\rm res})^2 \crm
& - \bar e_2^q \tilde H_1(\Phi,\Psi_{\rm res}) \cos\psi
}
\label{eq.Hamtilde2}
\ee
with
\be
\frac{1}{2}\left.\Dron{^2\Ham_0}{\Psi^2}\right|_{\Psi=\Psi_{\rm res}} = 
-\frac{3}{2}\frac{p^2 n_1}{\bar \Lambda_1}\frac{m_1+\bar \alpha m_2}{\bar
\alpha m_2}\ .
\label{eq.dHam2}
\ee
The Hamiltonian (\ref{eq.Hamtilde2}) is that of a pendulum. The
center of libration depends on the sign of $\tilde H_1(\Phi,\Psi_{\rm
res})$. If it is positive, as in most of the cases, the libration is
around $\psi=\psi_{\rm res}=\pi$, otherwise it is around $\psi_{\rm
res}=0$. Since the initial conditions used for the simulations in
section~\ref{sec.simu} ($\lambda_1=\lambda_2=\varpi_2=270\deg$) imply
$\psi(0)=0$, most of the resonances have been crossed at the hyperbolic
point. To study the TTV inside of a MMR, we thus change the initial
conditions, and take $\lambda_2=90\deg$. The other parameters are those
of the set 3 in the table~\ref{tab.param}.

\figcrossing

To get the evolution of $\phi(t)$, and then $\delta \lambda_1(t)$,
one should note that the oscillating term in $\dot \phi = \partial
\tilde \Ham / \partial \Phi$ (\ref{eq.dphi}) is proportional to
$(\Psi-\Psi_{\rm res})$ which is also proportional to $\dot \psi =
\partial \tilde \Ham / \partial \Psi$. One obtains
\be
\EQM{
\phi(t) =& \phi(0) -\frac{n_1 t}{p+q} \crm
&+\frac{3}{\bar \Lambda_1}\frac{p n_1}{p+q} \left(
\left.\Dron{^2 \tilde \Ham_0}{\Psi^2}\right|_{\Psi=\Psi_{\rm res}}
\right)^{-1} \big(\psi(t) - \psi(0)\big)\ .
}
\label{eq.phi2}
\ee
Removing the linear part of (\ref{eq.phi2}), and using
Eqs.~(\ref{eq.angle}), (\ref{eq.dHam2}), one finds
\be
\delta \lambda_1 = \frac{1}{p} 
\frac{\bar \alpha m_2}{m_1 + \bar \alpha m_2} \big(
\psi(t) - \psi_{\rm res})\ .
\ee
The corresponding TTV expression (\ref{eq.deltat}) is
\be
\delta t = \frac{1}{p n_1} \frac{\bar \alpha m_2}{m_1 + 
\bar \alpha m_2} \big( \psi(t) - \psi_{\rm res} \big)\ .
\label{eq.deltat2}
\ee
This expression is similar to that given by \citet{Agol_etal_MNRAS_2005}
for first order MMR. In (\ref{eq.deltat2}), the resonant angle $\psi$ is
librating. So, the absolute value of $(\psi(t)-\psi_{\rm res})$ is
bounded by the amplitude of libration, which is also bounded by $\pi$.

Figure~\ref{fig.crossing} shows the evolution of the dispersion $\sigma$
(\ref{eq.sigma}) of the TTV signal for systems in the vicinity the $3:7$
MMR.  The analytical approximation plotted as a solid curve is given by
(\ref{eq.deltat1}) outside of the MMR, and by (\ref{eq.deltat2}) inside
of the MMR, within $(a_2/a_{\rm res})^{3/2} \simeq 0.995$ and $1.004$.
In both cases, the analytical values are divided by $\sqrt{2}$ to
convert amplitudes into dispersions. Outside of the MMR, the dispersion
is larger for systems closer to the separatrices. This is due to the
factor $1/\varepsilon^2$ in (\ref{eq.deltat1}). Inside of the MMR, the
dispersion of TTV is also larger towards the separatrices, but the
reason is different \citep{Veras_etal_ApJ_2011}. Inside MMR, the
dispersion is proportional to the amplitude of libration which is
maximal at the separatrices.

For completeness, we provide also the libration frequency $\omega_{\rm
lib}$ at the center of resonant islands. This is an upper limit of the
frequency of the highest amplitude oscillation in TTV signals. Indeed,
closer to the separatrices, the period gets longer. The expression of
this frequency is
\be
\EQM{
\omega_{\rm lib} &= \sqrt{\abs{
\bar e_2^q \tilde H_1(\Phi,\Psi_{\rm res})
\left(
\left.\Dron{^2 \tilde \Ham_0}{\Psi^2}\right|_{\Psi=\Psi_{\rm res}}
\right)
}} \crm
&= p n_1 \sqrt{6 \left(\bar \alpha \frac{m_2}{m_0}+\frac{m_1}{m_0}\right)
\abs{C_1(\bar \alpha)} \bar e_2^q}\ .
}
\ee

\section{Conclusion}
A terrestrial planet at a period ratio larger than 2 with a Jovian
transiting planet may produce detectable TTV only if the system is at
resonance, but not in the exact center, or near period commensurability.
If a system is sufficiently close to a second or higher order resonance,
to be detected, then the TTV signal is dominated by a sinusoidal
oscillation which \bfx{is difficult to distinguish} from one resonance to an
other. Moreover, the derived analytical approximations of TTV signals
Eqs.~(\ref{eq.deltat1}) and (\ref{eq.deltat2}) show that once the main
period $P$ of the TTV is known, the amplitude still depends on the mass
$m_2$, the eccentricity $e_2$, and the period $P_2$ (through the
semi-major axis ratio $\bar \alpha$) of the perturber. The problem is
thus highly degenerate. The examples presented in this letter show that
an Earth mass planet is able to produce a signal comparable to that
produced by a Saturn mass planet. We thus conclude that the TTV method
may benefit from radial velocity observations in order to characterize
non-transiting planets.

\section*{acknowledgements}
This work was supported by the European Research Council/European
Community under the FP7 through Starting Grant agreement number 239953.
We also acknowledge the support from Funda\c{c}\~ao para a Ci\^encia e
a Tecnologia (FCT) through program Ci\^encia\,2007 funded by FCT/MCTES
(Portugal) and POPH/FSE (EC), and in the form of grant reference
PTDC/CTE-AST/098528/2008.

\bibliographystyle{mn2e}
\bibliography{ttv}

\bsp

\label{lastpage}

\end{document}